\documentclass{bioinfo}
\copyrightyear{2005}
\pubyear{2005}

\begin{document}
\firstpage{1}

\title[Modeling biological networks]{An integrative approach to modeling biological networks}
\author[Memi\v{s}evi\'{c} \textit{et~al}]{Vesna Memi\v{s}evi\'{c}, Tijana Milenkovi\'{c}, and Nata\v{s}a Pr\v{z}ulj\,$^{*}$}

\address{$^{}$Department of Computer Science, University of California, Irvine, CA 92697-3435, USA \\
$^{*}$Corresponding author (e-mail: natasha@ics.uci.edu)}

\maketitle
\medskip

\section*{Abstract}
\textbf{Background:}
Networks are used to model real-world phenomena in various domains,
including systems biology.  Since proteins carry out biological
processes by interacting with other proteins, it is expected that
cellular functions are reflected in the structure of
protein-protein interaction (PPI) networks. Similarly, the topology of
residue interaction graphs (RIGs) that model proteins' 3-dimensional
structure might provide insights into protein folding, stability, and
function. An important step towards understanding these networks is
finding an adequate network model.  Evaluating the fit of a model
network to the data is a formidable challenge, since network
comparisons are computationally infeasible and thus rely on
heuristics, or ``network properties.''

\textbf{Results:}
We show that it is difficult to assess the reliability of the fit of a
model using any network property alone. Thus, we present an
integrative approach that feeds a variety of network properties into
five probabilistic methods to predict the best-fitting network model
for PPI networks and RIGs.  We confirm that geometric random graphs
(GEO) are the best-fitting model for RIGs.  Since GEO networks model
spatial relationships between objects and are thus expected to
replicate well the underlying structure of spatially packed residues
in a protein, the good fit of GEO to RIGs validates our approach.
Additionally, we apply our approach to PPI networks and confirm that
the structure of merged data sets containing both binary and
co-complex data that are of high coverage and confidence is also
consistent with the structure of GEO, while the structure of sparser
and lower confidence data is not.  Since PPI data are noisy, we test
the robustness of the five classifiers to noise and show that their
robustness levels differ.

\textbf{Conclusions:}
We demonstrate that none of the classifiers predicts noisy scale-free
(SF) networks as GEO, whereas noisy GEOs can be classified as SF.
Thus, it is unlikely that our approach would predict a real-world
network as GEO if it had a noisy SF structure.  However, it could
classify the data as SF if it had a noisy GEO structure.  Therefore,
the structure of the PPI networks is the most consistent with the
structure of a noisy GEO.

\medskip

\textbf{Keywords:}   network modeling, biological networks, protein-protein interaction networks, residue interaction graphs

\medskip

\section*{Background}

Large-scale biological network data are increasingly becoming
available due to advances in experimental biology. We analyze
protein-protein interaction (PPI) networks, where proteins are modeled
as network nodes and interactions amongst them as network edges. Since
it is the proteins that carry out almost all biological processes and
they do so by interacting with other proteins, analyzing PPI network
structure could lead to new knowledge about complex biological
mechanisms and disease.  Additionally, we analyze network
representations of protein structures, ``residue interaction graphs''
(RIGs), where residues are modeled as network nodes and inter-residue
interactions as network edges; an inter-residue interaction exists
between residues that are in close spatial proximity. Understanding
RIGs might provide deeper insights into protein structure, binding,
and folding mechanisms, as well as into protein stability and
function.

To understand these complex biological network data, one must be able
to successfully reproduce them. Finding an adequate network model that
will generate networks that closely replicate the structure of real
data is one of the first steps in this direction. Thus, we focus on
finding well-fitting network models for biological networks. The hope
is that a good network model could provide insights into understanding
of biological function, disease, and evolution. For example, the use
of an adequate network model is vital for discovering network motifs,
evolutionary conserved functional modules
\cite{Shen-Orr02,Milo02,Milo04}, since network motif discovery
requires comparing real-world networks with randomized ones
\cite{Milo04}. A well-fitting network model could also be used to
assign confidence levels to existing protein interactions, as well as
to predict new interactions that were overlooked experimentally
\cite{Kuchaiev2009}. Additionally, it could guide biological
experiments in a time- and cost-optimal way, thus minimizing the costs
of interactome detection \cite{LappeHolm04}.  Since discovering PPI
and other biological networks is in its infancy, it is expected that
practical application of network models will increase and prove its
value in the future.

\subsection*{Overview}

Several network models have been proposed for biological
networks. Starting with Erd\"{o}s-R\'{e}nyi random graphs
\cite{ErdosRenyi59}, various network models have been designed to
match certain properties of real-world networks. Early studies
published largely incomplete yeast two-hybrid PPI data sets
\cite{Ito00,uetz00} that were well modeled by scale-free networks
\cite{Barabasi_Oltvai04,Jeong01}. In a scale-free network, the
distribution of degrees follows a power-law \cite{Barabasi99}.
Modeling of the data by scale-free networks was based on the
assumption that the degree distribution is one of the most important
network parameters that a good network model should capture.  However,
networks of vastly different structures could have the same degree
distributions \cite{Tanaka05a}.  Additionally, it has been argued that
currently available PPI network data are samples of the full
interactomes and thus the observed power-law degree distributions are
artifacts of sampling properties of these networks
\cite{Stumpf05,Vidal05,deSilva2006}.  As new biological network data
becomes available, we need to ensure that our models continue to fit
the data well.  In the light of new PPI network data, several studies
have started questioning the wellness of fit of scale-free models: an
evidence has been presented that the structure of PPI networks is
closer to geometric random graphs, that model spatial relationships
between objects, than to scale-free networks
\cite{Przulj04,Przulj06,HRP08}. Similarly, geometric
random graph model has been identified as an optimal network model for
RIGs \cite{Milenkovic2008}.

\begin{figure*}[!ht]
\begin{center}
\begin{tabular}{cc}
\textbf{(a)}{\resizebox{0.45\textwidth}{!}{\includegraphics{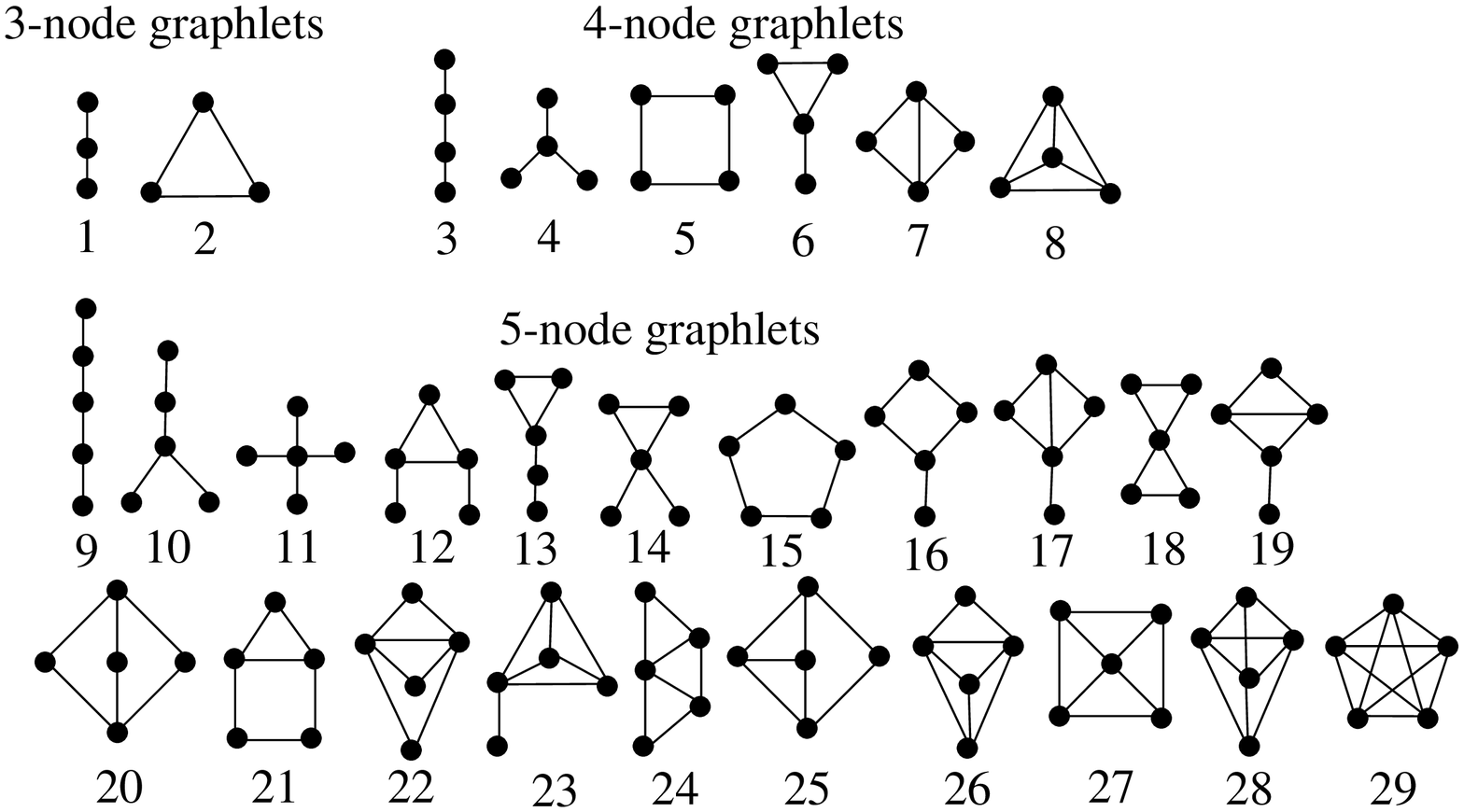}}}
\hspace{1cm}
\textbf{(b)}{\resizebox{0.33\textwidth}{!}{\includegraphics{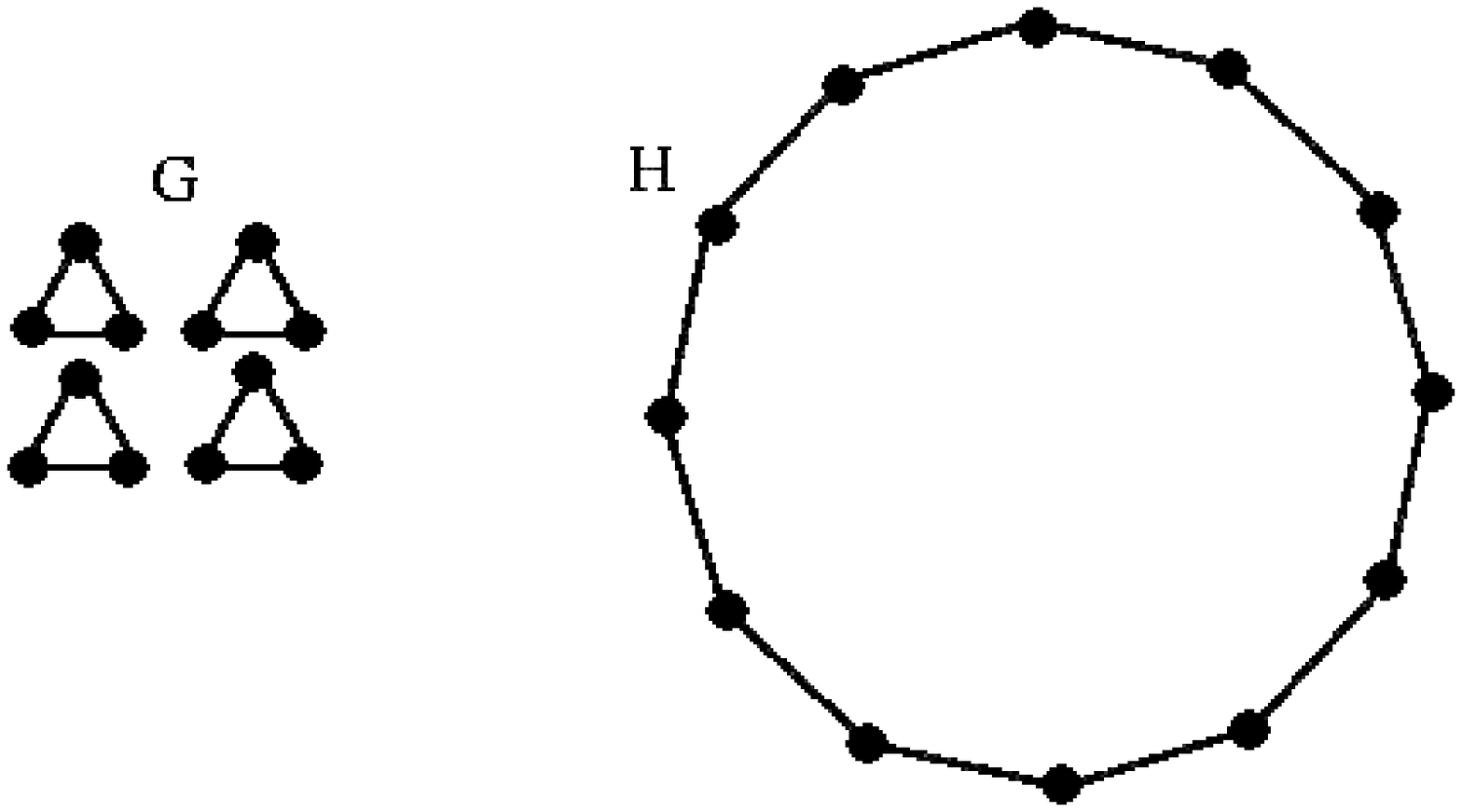}}}
\end{tabular}
\end{center}
\caption{\textbf{(a)} All 3-node, 4-node and 5-node graphlets \citep{Przulj04}; \textbf{(b)} an example of two networks of the same size, $G$ and $H$, that have the same degree distribution, but very different network structure.}
\label{fig:heur}
\end{figure*}

A well-fitting network model should generate graphs that closely
resemble the structure of real-world networks. To evaluate the fit of
a model to the data, one needs to compare model networks with
real-world networks. However, network comparisons are computationally
infeasible due to NP-completeness of the underlying subgraph
isomorphism problem \cite{Cook1971}. Therefore, large network
comparisons rely on heuristics, commonly called ``network
properties.'' These properties belong to two major classes: global and
local.  Global properties include the degree distribution, average
clustering coefficient, clustering spectrum, average diameter, and the
spectrum of shortest path lengths. Local properties include network
motifs, small overrepresented subgraphs \cite{Milo02,Milo04}, and
graphlets, small connected induced subgraphs of real-world networks
(Figure \ref{fig:heur} (a)) \cite{Przulj04}. Based on graphlets, two highly sensitive
measures of network local structural similarities were designed: the
relative graphlet frequency distance (``RGF-distance'')
\cite{Przulj04} and graphlet-based generalization of the degree
distribution, called graphlet degree distribution agreement
(``GDD-agreement'') \cite{Przulj06}.  The choice of a network property
for evaluating the fit of a network model to the data is non-trivial,
since different models might be identified as optimal with respect to
different properties. In general, global properties might not be
constraining enough to capture complex topological characteristics of
biological networks. For example, two networks with exactly the same
degree distributions can have completely different underlying
topologies (Figure \ref{fig:heur} (b)). On the other hand, local properties,
RGF-distance and GDD-agreement, impose a larger number of constraints,
thus reducing degrees of freedom in which networks being compared can
differ.  The fit of model networks to real-world data can also be
evaluated by using principal component analysis of the vector space
whose coordinates are the statistics of
network properties \cite{Filkov2009}, as well as by counting the
number of random walks of a given length in the network and feeding
these counts into a probabilistic method
\cite{Middendorf2004,Middendorf2005}.
%

\subsection*{Our Contribution}

Since it might be difficult to assess the reliability of any
particular model with any one statistic (which we demonstrate below),
we introduce a novel approach for evaluating the fit of network models
to real-world networks.  Our approach integrates a variety of local
and global network properties into the ``network fingerprint,'' a
vector whose coordinates contain the following network properties: the
average degree, the average clustering coefficient, the average
diameter, and the frequencies of appearance of all 31 graphlets with 1
to 5 nodes (see Methods for details).  Additionally, our approach
applies a series of probabilistic methods (also called classifiers) to
network fingerprints to predict the best-fitting network model.

Our method proceeds through the following steps. First, we represent
each real-world and model network with its fingerprint.  Second, we
use fingerprints of model networks as input into probabilistic methods
to train them. Third, we validate the prediction accuracy of each
probabilistic method. Next, we use network fingerprints of real-world
networks as input into trained probabilistic methods to predict their
best-fitting network models.  Finally, we provide several validations
of our model predictions.

\section*{Results and Discussion}

\subsection*{Data Sets}

We need to distinguish between two different types of PPIs: binary
interactions obtained by yeast two-hybrid (Y2H) technique and
co-complex data obtained by mass spectrometry of purified complexes.
Since in co-complex data interactions are defined by using either the
``spoke'' or the ``matrix'' model, binary interaction networks are
believed to have fewer false positives than co-complex data
\cite{Venkatesan2009,Simonis2009}; in the spoke model, edges exist
between the bait and each of the preys in a pull-down experiment, but
not between the preys, while in the matrix model, additional edges are
formed between all preys. However, due to technological
limitations of Y2H, binary interaction networks still contain many
false negatives and are thus incomplete
\cite{Venkatesan2009,Simonis2009}.  Networks from large databases
contain both binary and co-complex PPIs; this makes them more
complete, but at the same time, they have high levels of false
positives. Also, they contain a large fraction of interactions
obtained by error-prone literature curation; since most of these
interactions are supported by a single publication, they are of
questionable quality
\cite{Cusick2009}.

We analyze physical PPI networks of four eukaryotic organisms: yeast
\emph{Saccharomyces cerevisiae}, fruitfly \emph{Drosophila
melanogaster}, worm \emph{Caenorhabditis elegans}, and human
\emph{Homo sapiens}.  We analyze the total of 12 PPI networks, 5 of
which are yeast, 3 of which are fruitfly, 1 of which is worm, and 3 of
which are human.

We denote PPI networks as follows.  ``YH1'' and ``YE1'' are the
high-confidence and the entire yeast PPI networks by Collins et
al. \cite{Krogan2007}, respectively. ``YH2'' is the yeast high
confidence PPI network described by von Mering et al. \cite{Mering02}.
``YE2'' is the yeast PPI network containing top 11,000 high-, medium-,
and low-confidence interactions from the same study
\cite{Mering02}. ``YE3'' is the entire physical yeast protein
interaction network from BioGRID \cite{BIOGRID}.  ``FH1'' and ``FE1''
are the high-confidence and the entire fruitfly PPI networks by Giot
et al. \cite{GiotSci03}, respectively. ``FE2'' is the entire physical
fruitfly protein interaction network from BioGRID
\cite{BIOGRID}. ``WE1'' is the entire worm PPI network from BioGRID
\cite{BIOGRID}.  Finally,
``HE1'' is the entire human PPI network by Rual et al. \cite{Rual05},
while ``HE2'' and ``HE3'' are entire human PPI networks from BioGRID
\cite{BIOGRID} and HPRD \cite{hprd}, respectively.
All five yeast PPI networks, as well as FE2, WE1, HE2, and HE3,
contain both binary and co-complex data. The remaining networks, i.e.,
FH1, FE1, and HE1, contain solely binary interactions.

In addition to PPI networks, we apply our approach to network
representations of protein structures, residue interaction graphs
(RIGs). In RIGs, nodes represent amino acids and edges exist between
residues that are close in space.  We analyze RIGs constructed for
nine structurally and functionally different proteins
\cite{Milenkovic2008}.  For each of the nine proteins, multiple RIGs
are constructed as undirected and unweighted graphs, with residues
\emph{i} and \emph{j} interacting if any heavy atom of residue
\emph{i} is within a given distance cut-off of any heavy atom of
residue \emph{j}. Various distance cut-offs in [4.0, 9.0] \AA\ are
used, as well as three different representations of residues: (1) RIGs
that contain as edges only residue pairs that have heavy
\emph{backbone} atoms within a given distance cut-off (``BB''), (2)
RIGs that contain as edges only residue pairs that have heavy
\emph{side-chain} atoms within a given distance cut-off (``SC''), and
(3) the most commonly used RIG model, in which \emph{all} heavy atoms
of every residue are taken into account when determining residue
interactions (``ALL''). In total, these different RIG definitions
result in 513 RIGs corresponding to nine different proteins (see
\cite{Milenkovic2008} for details).

\subsection*{Techniques}

We apply five commonly used probabilistic methods: backpropagation
method (``BP''), probabilistic neural networks (``PNN''), decision
tree (``DT''), multinomial na\"{i}ve Bayes classifier (``MNB''), and
support vector machine (``SVM'') (see Methods). We evaluate the fit of
real-world networks to three different network models:
Erd\"{o}s-R\'{e}nyi random graphs (``ER'') \cite{ErdosRenyi59},
preferential attachment scale-free networks (``SF'')
\cite{Barabasi99}, and 3-dimensional geometric random graphs (``GEO'')
\cite{Penrose03, Przulj04} (see Methods). We do not consider other
commonly used network models, such as random graphs with the same
degree distribution as the data \cite{Mol95}, or the stickiness
index-based network model \cite{PrzuljHigham06}; generating these
models requires as input the degree distribution of real-world
networks, while the training and testing sets of random networks need
to be generated without any data input.

We start by generating the set of 8,220 random networks of different
sizes belonging to the three network models: ER, SF, and GEO (see
Methods). We divide these random networks into two sets: the
``training set,'' containing 20\% of them, and the ``testing set,''
containing the remaining 80\% of them.  We choose this ratio for the
training and the testing sets to achieve good training and
generalization of probabilistic methods, as well as to avoid data
over-fitting. Next, we find fingerprints for these model networks and
provide them as input into probabilistic classifiers. We train the
five probabilistic methods on random networks from the training set,
so that probabilistic classifiers could learn to distinguish between
fingerprints of random networks belonging to different models. Then,
we validate prediction accuracies of probabilistic methods on the
testing set. That is, we examine how well probabilistic methods work
on new, yet unseen data, by analyzing whether they classify random
networks from the testing set into their correct models. We define the
validation rate of a probabilistic method as the percentage of random
networks from the testing set that are correctly classified.
Thus, the validation rate can be interpreted as the likelihood that
a probabilistic method will classify a network to its correct model.
The validation rates over the entire testing data set for BP, PNN, DT,
MNB, and SVM are 99.98\%, 99.97\%, 99.41\%, 98.48\%, and 94.72\%,
respectively (column 2 of \ref{tab:1} 1).  Model-specific validation rates
are presented in columns 3-5 of Table \ref{tab:1}.  These high validation rates
indicate that all five probabilistic classifiers are able to
successfully classify random networks into their correct models.  We
also verify that the probabilistic methods are robust to noise (see
below), which is important since we are dealing with noisy PPI data.
For these reasons, we believe that our approach correctly classifies
biological networks into their best-fitting network models.

\begin{table}[!ht]
{\begin{tabular}{p{0.6in}||p{0.6in}|p{0.6in}|p{0.6in}|p{0.6in}|}
\hline Classifier & VR-Total   &  VR-ER & VR-GEO & VR-SF \\
\hline
\hline BP & 99.98\% (6,575/6,576) & 100\% (2,192/2,192) & 100\%    (2,192/2,192) & 99.96\% (2,191/2,192) \\
\hline PNN & 99.97\%    (6,574/6,576) & 100\% (2,192/2,192) & 100\% (2,192/2,192) & 99.91\%    (2,190/2,192) \\
\hline DT & 99.41\% (6,537/6,576) & 99.41\%    (2,179/2,192) & 99.64\% (2,184/2,192) & 99.18\% (2,174/2,192) \\
\hline MNB & 98.48\% (6,476/6,576) & 98.18\% (2,152/2,192) & 100\%  (2,192/2,192) & 97.26\% (2,132/2,192) \\
\hline SVM & 94.72\%    (6,229/6,576) & 94.85\% (2,079/2,192) & 100\% (2,192/2,192) &    89.33\% (1,958/2,192) \\
\hline
\end{tabular}
}\\
\bigskip
\caption{The validation rates (``VR'') for the five probabilistic classifiers,
BP, PNN, DT, MNB, and SVM (column 1), over the entire testing set of
6,576 ER, GEO, and SF networks (column 2), as well as within
each individual testing subset of 2,192 ER, 2,192 GEO, or 2,192 SF
networks (columns 3--5, respectively}
\label{tab:1}
\end{table} 

\subsection*{Results}

\begin{figure}[ht]
        \centering \scalebox{.45}{\includegraphics{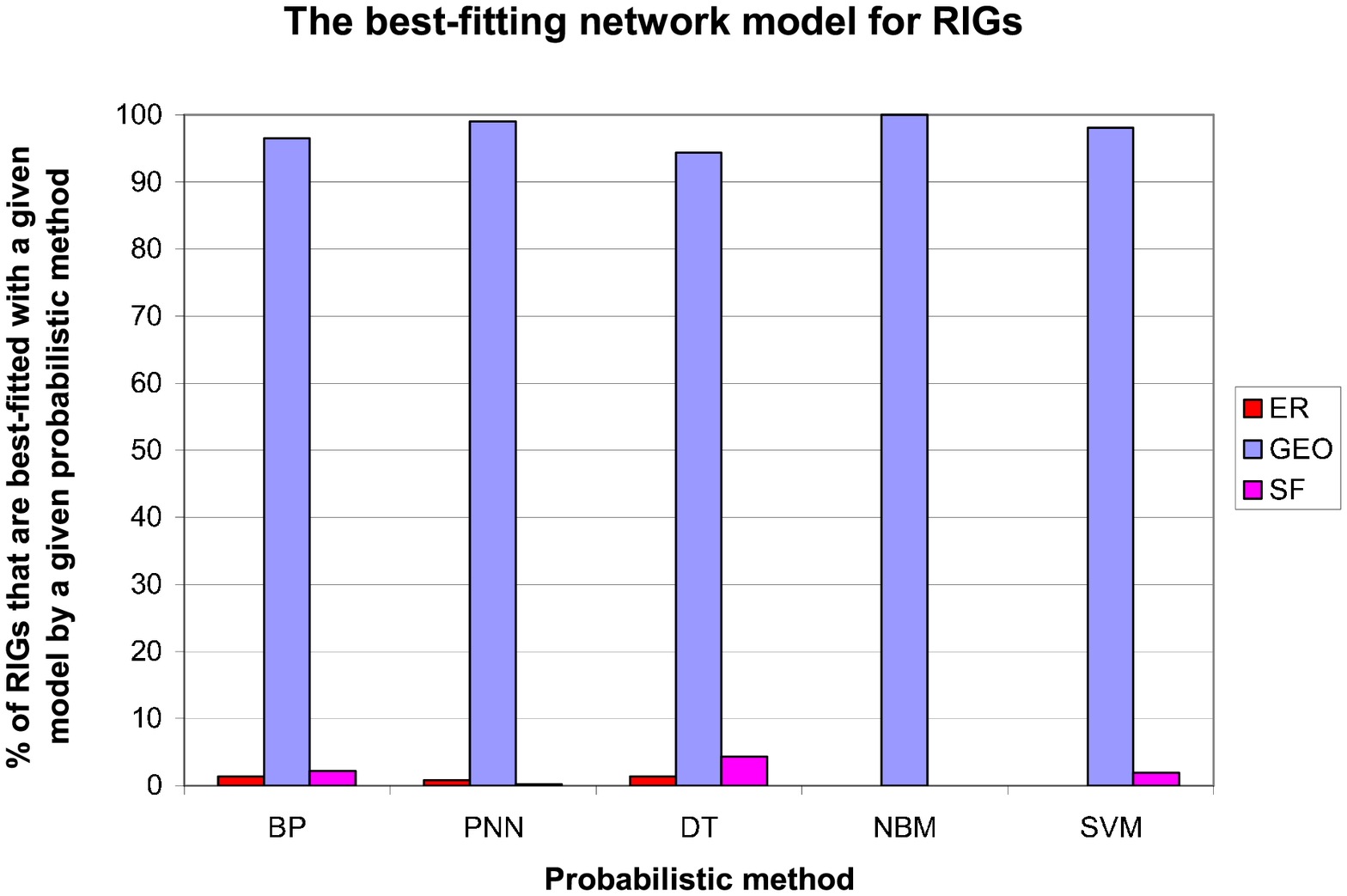}}
        \caption{The best-fitting network model out of the three models (ER, GEO, and SF) predicted by the five probabilistic classifiers (BP, PNN, DT, NBM, and SVM) for the 513 analyzed RIGs.}
        \label{fig:rigs_embedding}
\end{figure}

The best-fitting network models for RIGs identified by each of the
five probabilistic methods are presented in Figure \ref{fig:rigs_embedding}. For more than
94\% of all analyzed RIGs, all five probabilistic methods predict GEO
as the best-fitting network model. This result is encouraging, since
GEO models spatial relationships between objects, and therefore, it is
expected to replicate well the underlying nature of spatially packed
residues in a protein. Our result is consistent with a recent study
that demonstrated, by using a variety of individual network
properties, that GEO is the optimal network model for RIGs
\cite{Milenkovic2008}.  The RIGs that are better modeled by SF and ER
networks are those that were constructed by using the lowest distance
cut-offs for ``SC'' contact type and the highest distance cut-offs for
``ALL'' contact type (see Section ``Data Sets'').  This is consistent
with our previous results \cite{Milenkovic2008}, therefore
additionally validating the correctness of this study.

Next, we apply our approach to PPI networks, which are, unlike RIGs,
noisy and incomplete, and therefore the identification of their
optimal network model could be more challenging. The best-fitting
network models for PPI networks predicted by each of the five
probabilistic methods are presented in Table 2.  Probabilistic
classifiers predict GEO as the best-fitting network model for most of
the analyzed yeast PPI networks: YH1, YE1, YH2, and YE2 (Table \ref{tab:2}).
This is encouraging, since yeast has the most complete interactome
\cite{Hart2006}, as indicated by high edge densities and clustering
coefficients of its PPI networks (Table \ref{tab:2}).  Additionally, yeast PPI
networks that are fitted the best by GEO are of high confidence, since
they were obtained by merging and de-noising multiple PPI data sets
\cite{Krogan2007,Mering02}. For example, YH1 is currently of the
highest confidence: it is comparable to small-scale experiments by the
quality of its interactions \cite{Krogan2007}.  For this network, all
five probabilistic methods predict GEO as the best-fitting network
model. These results are consistent with studies that demonstrated the
superiority of the fit of GEO to newer, more complete and less noisy
PPI networks \cite{Przulj04,Przulj06,HRP08}.

Out of the remaining PPI networks in Table \ref{tab:2}, three are binary
interaction data sets (FH1, FE1, and HE1), and five originate from
large PPI databases, BioGRID and HPRD \cite{BIOGRID,hprd}, that
contain both binary and co-complex data (YE3, FE2, WE1, HE2, and HE3).
Binary PPI networks are less noisy, but also less complete, while
networks from large databases have higher coverage, but are more
noisy, as described in Section ``Data Sets''
\cite{Venkatesan2009,Simonis2009}. Additionally, large databases
contain a large fraction of interactions obtained by literature
curation (LC) \cite{BIOGRID}.  It has recently been shown that LC can
be error-prone and possibly of lower quality than commonly believed
\cite{Cusick2009,Venkatesan2009}.  Given that more than 75\% (85\%) of
the LC yeast (human) PPIs in BioGRID are supported by a single
publication \cite{Cusick2009}, the quality of these interactions might be
questionable \cite{Venkatesan2009}. Moreover, a considerably low
overlap between high-throughput experimental and LC PPIs in BioGRID
\cite{BIOGRID}, as well as a surprisingly low overlap of
interactions across different databases \cite{Cusick2009}, might
suggest that many interactions still remain to be validated and
discovered \cite{BIOGRID,Cusick2009,Venkatesan2009}. For these
reasons, it is not surprising that SF and ER are the best-fitting
models for binary Y2H PPI networks and for PPI networks from large
databases (Table \ref{tab:2}). Since PPI networks are unlikely to be organized
completely at random, the best fit of ER to some of them additionally
verifies the presence of noise. A good fit of SF to networks that are
smaller samples of complete interactomes (obtained only by Y2H) is
consistent with previous studies arguing that power-law degree
distributions in PPI networks are an artefact of their sampling
\cite{Stumpf05,Vidal05,deSilva2006}.

\subsubsection*{Robustness and Validation.}

To test the robustness of our approach to noise, we randomly add,
remove, and rewire 10\%, 20\%, and 30\% of edges in YH1 network and
its corresponding model networks and examine how the probabilistic
methods classify them (Table \ref{tab:random}). We test the robustness on YH1, since
this network is of the highest confidence \cite{Krogan2007}. Clearly,
there is no need to introduce noise in ER networks, since they cannot
be made more random. It is expected that with the introduction of more
noise of ER type into the data and model networks, noisier networks
will increasingly be classified as ER.  Indeed, SVM classifies SF
networks with 20-30\% of edges deleted and rewired as ER (Table \ref{tab:random}). At
lower levels of noise, all classifiers predict noisy SF to still be SF
(Table \ref{tab:random}).  Thus, noisy SF (and clearly, ER) are never classified as
GEO.  Similarly, increasing levels of noise in GEO networks cause
their increasing miss-classification into ER or SF models.  Thus,
noisy GEO can be classified as either GEO, SF, or ER. This
demonstrates that our approach is unlikely to classify a real-world
network that has a noisy SF or ER topology as GEO. On the other hand,
it might classify a real-world network that has a noisy GEO topology
either as GEO, SF, or ER. Thus, the yeast PPI networks that are
classified as GEO are unlikely to have SF or ER network
structure. However, PPI networks of any organism that are classified
as SF or ER could have noisy GEO structure.

The probabilistic method that is the most robust to noise is MNB,
since it always correctly predicts the model irrespective of the level
of noise (Table \ref{tab:random}). The least robust method seems to be DT, since it
always predicts noisy GEO networks as SF or ER.  Note however, that
this is not surprising, since small changes in the input of a decision
tree may cause large changes in its output due to a relative
sensitivity of branching to the input values. For this reason, it is
not surprising that DT incorrectly classifies YH2 and YE2 networks
that are predicted to be GEO by most other classifiers (Table \ref{tab:2}).

\begin{table}[!ht]
{\begin{tabular}{|p{0.18in}||p{0.23in}|p{0.28in}|p{0.2in}|p{0.19in}|p{0.19in}|p{0.19in}|p{0.19in}|p{0.19in}|p{0.19in}|p{0.19in}| }
\hline
Data & \# of nodes & \# of edges & Avg diam & Avg cc & BP & PNN & DT & NBM & SVM\\
\hline
\hline
YH1 & 1,622 &  9,074 & 5.53 & 0.55  & GEO & GEO & GEO & GEO & GEO\\
\hline
YE1 & 2,390 & 16,127 & 4.82 & 0.44  &  GEO & ER & GEO & GEO & GEO\\
\hline
YH2 & 988 & 2,455 & 5.19 & 0.34 &  GEO & GEO & SF & GEO & GEO\\
\hline
YE2 & 2,401 & 11,000 & 4.93 & 0.30 &  GEO & ER & SF & GEO & GEO\\
\hline
YE3 & 4,961  & 39,434 & 3.48 & 0.18  &  SF & ER & SF & SF & ER\\
\hline
FH1 & 4,602 & 4,637 & 9.44 & 0.02  &  SF & ER & ER & SF & ER\\
\hline
FE1 & 6,985 & 20,007 & 4.47 & 0.01  &  SF & SF & SF & SF & SF\\
\hline
FE2 & 7,040  & 22,265 & 4.34 & 0.01  &  SF & SF & SF & SF & SF\\
\hline
WE1 & 3,524 & 6,541 & 4.32 & 0.05 &  SF & SF & SF & SF & SF\\
\hline
HE1 & 1,873 & 3,463 & 4.34 & 0.03  &  SF & SF & SF & SF & ER\\
\hline
HE2 & 7,941 & 23,555 & 4.69 & 0.11  &  SF & SF & SF & SF & SF\\
\hline
HE3 & 9,182 & 34,119 & 4.26 & 0.10  &  SF & SF & SF & SF & SF\\
\hline
\end{tabular}}\\
\bigskip
\caption{The best-fitting network models (out of ER, GEO, and SF) predicted by the five probabilistic classifiers (BP,
PNN, DT, NBM, and SVM) for the 12 PPI networks.  The PPI networks are presented in the first column, denoted by ``Data.''  Columns two to five contain the number of nodes, the number of edges, the average diameter, and the average clustering coefficient of a network, respectively.  Columns six to ten contain network models predicted by the five classifiers for each of the PPI networks.}
\label{tab:2}
\end{table} 

\begin{table}[!ht]
{\resizebox{0.482\textwidth}{!}{\includegraphics{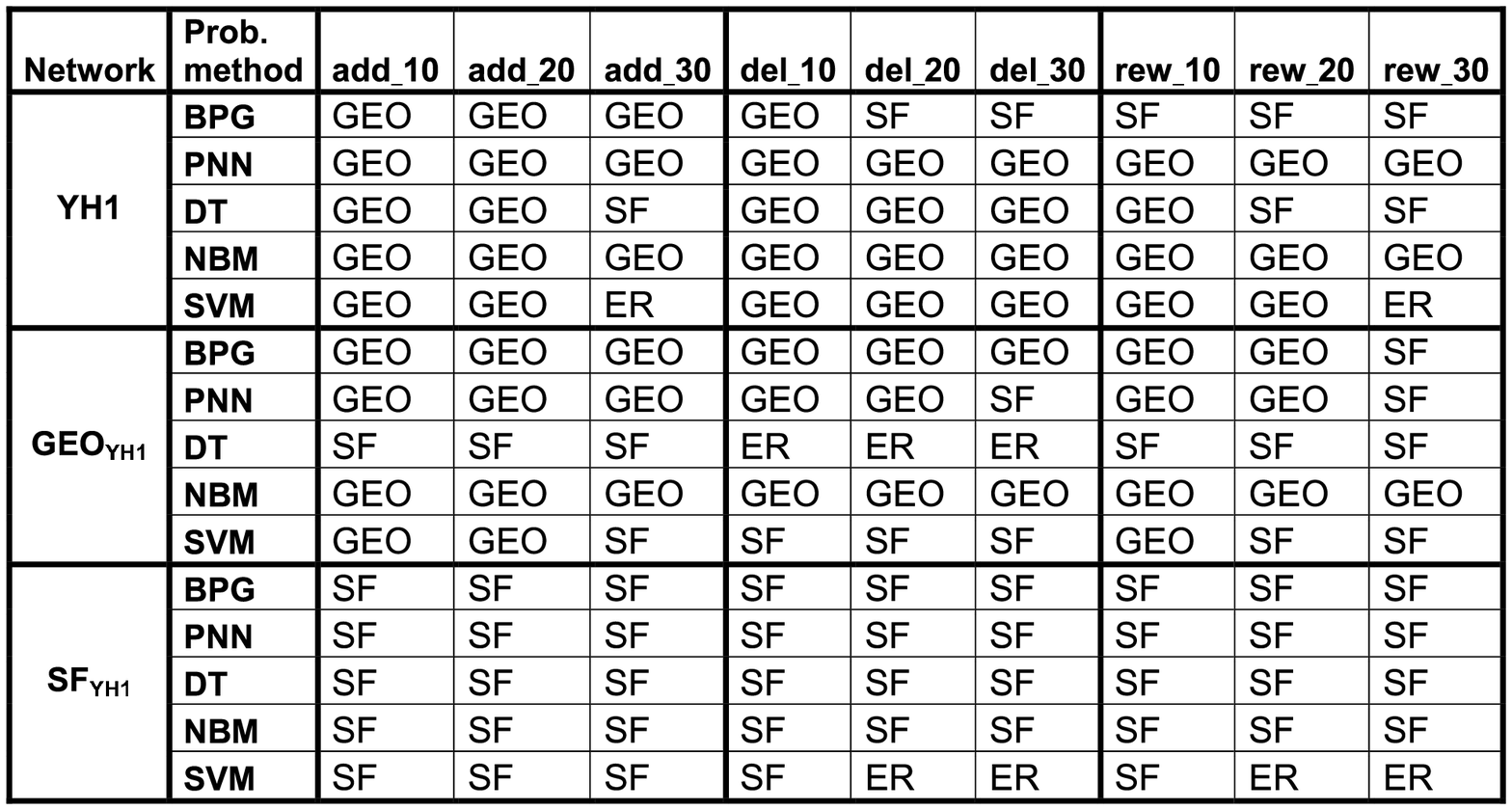}}}
\caption{The best-fitting network models (ER, GEO, SF) predicted by the five probabilistic classifiers (BP,
PNN, DT, NBM, and SVM) for noisy networks.  The networks to which the
noise is added are: YH1 network, as well as a GEO and an SF network of
the same size as YH1, denoted by ``GEO$_{\textnormal{YH1}}$'' and
``SF$_{\textnormal{YH1}}$'', respectively (listed in column 1).
We obtained noisy networks by randomly adding, deleting, and rewiring
10\%, 20\%, and 30\% of edges (columns 3-11, respectively).  For each
of YH1, GEO$_{\textnormal{YH1}}$ and SF$_{\textnormal{YH1}}$ and for
each of the randomization schemes, we constructed 10 instances of
noisy (randomized) networks, resulting in the total of $3 \times 9
\times 10=270$ noisy networks.  For each of YH1, GEO$_{\textnormal{YH1}}$ and
SF$_{\textnormal{YH1}}$, the classifiers predicted the same model for all instances
of noisy networks in the same randomization scheme; predicted models are reported in columns
3-11.}
\label{tab:random} \end{table}

We take a step further towards validating our results. We apply an
algorithm that directly tests whether PPI networks have a geometric
structure by embedding the proteins into a low-dimensional space given
only their PPI network connectivity information \cite{HRP08}.  We
embed in 3-dimensional (3D) Euclidian space, simply as a proof of
concept. The algorithm is based on multidimensional scaling
\cite{MDS}, with shortest path lengths between protein pairs in a PPI
network playing the role of Euclidean distances in space. After
proteins are embedded in space, a radius $r$ is chosen so that each
node is connected to the nodes that are at most at distance $r$ from
it; this procedure results in construction of a geometric graph (as
defined in Methods section below). Each choice of a radius thus
corresponds to a different geometric graph. By varying the radius,
specificity and sensitivity are measured to quantify the ability of
each constructed geometric graph to recover the original PPI
network. Then, the overall goodness of fit is judged by computing the
areas under the Receiver Operator Characteristic (ROC) curves, with
higher values indicating a better fit \cite{Bradley1997}. For details,
see \cite{HRP08}.

\begin{figure*}[!ht]
\begin{center}
\begin{tabular}{cc}
\textbf{(a)}{\resizebox{0.461\textwidth}{!}{\includegraphics{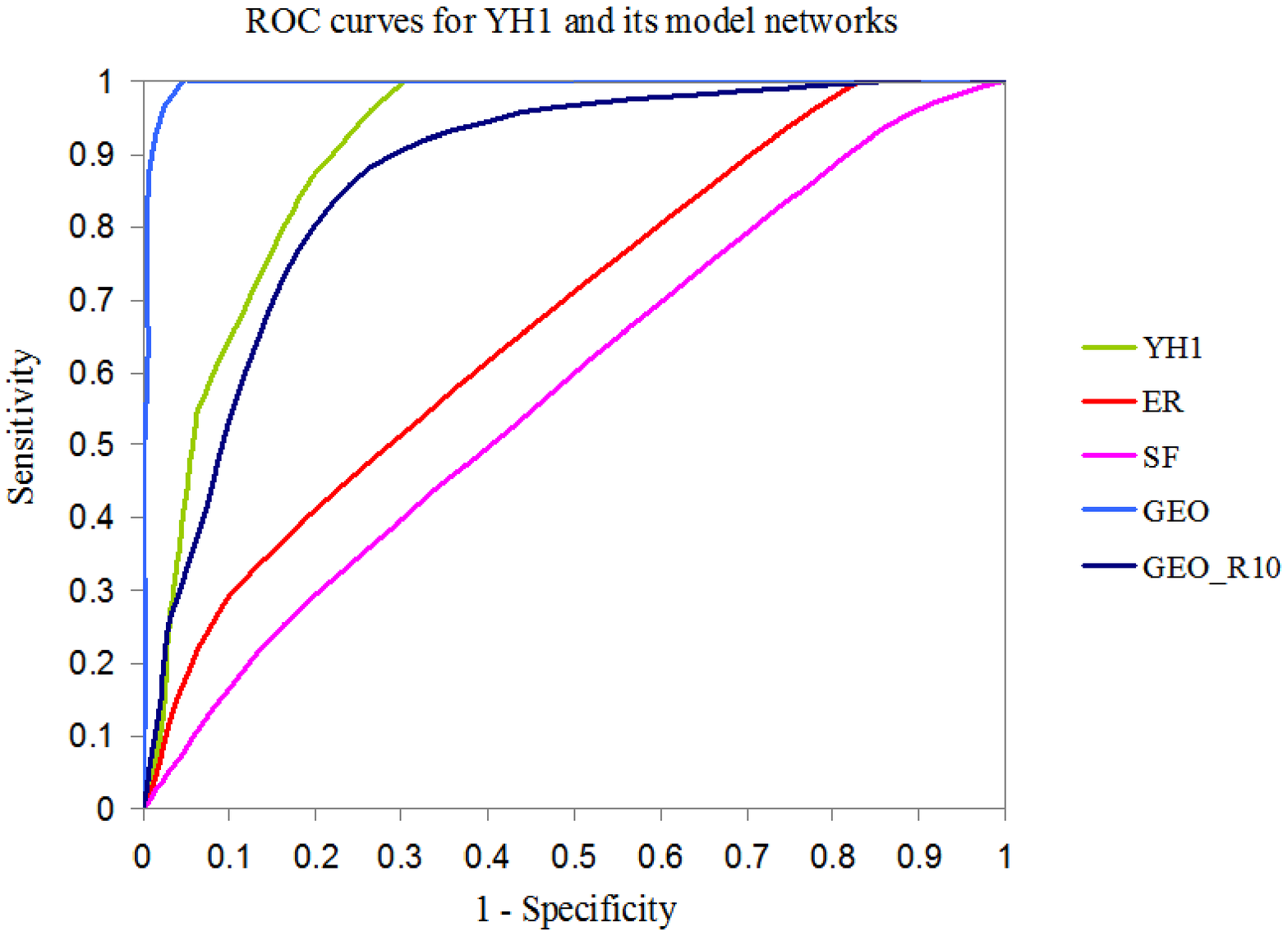}}}
\textbf{(b)}{\resizebox{0.485\textwidth}{!}{\includegraphics{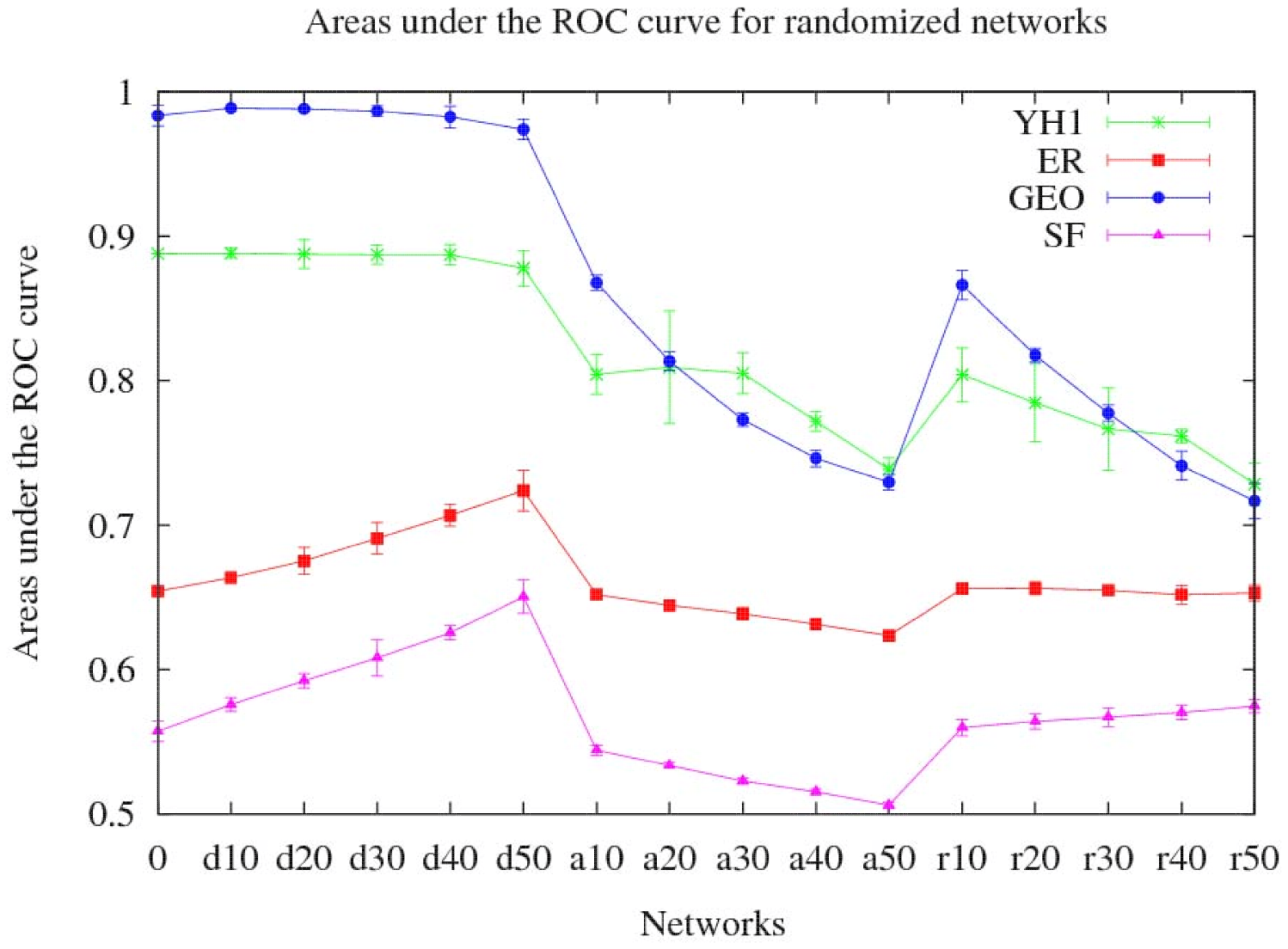}}}
\end{tabular}
\end{center}
\caption{\textbf{(a)} ROC curves illustrating the performance of the embedding algorithm for YH1 PPI network and one network belonging to each of the following model networks that are of the same size as YH1: ER, SF, GEO, and randomized GEO network (``GEO\_R10'') obtained by randomly rewiring 10\% of edges. \textbf{(b)} Areas under the ROC curve (AUCs) for YH1 and its ER, GEO, and SF model networks (denoted by ``0'' on x-axis), as well as for their randomized versions
obtained by randomly deleting, adding, and rewiring (denoted by ``d'', ``a'', and ``r'' on x-axis, respectively) 10\%, 20\%, 30\%, 40\%, and 50\% of their edges (denoted by ``10'', ``20'', ``30'', ``40'', and ``50'' on x-axis, respectively). For each of the network models and each of the randomization schemes, points in the panel represent
averages of AUCs over 10 networks. The error bar around a point is one standard deviation below and above the point.}
\label{fig:rocs_model}
\end{figure*}

We apply this algorithm to YH1 PPI network, as well as to ER, GEO, and
SF model networks of the same size as YH1.  As expected, the resulting
areas under the ROC curve (AUCs) are low for ER and SF, with values of
0.65 and 0.56, respectively, since these networks do not have a
geometric structure (Figure \ref{fig:rocs_model} (a)).  On the other hand, AUCs are high
for the data and GEO, with values of 0.89 and 0.98, respectively,
suggesting that the data has a geometric structure (Figure \ref{fig:rocs_model} (a)).  For
each of the network models, the reported AUC is the average over 10
random graphs.

Since PPI networks are noisy, we test how robust the embedding
algorithm is to noise in the data and model networks. We add noise to
YH1 and its corresponding ER, SF, and GEO model networks by randomly
deleting, adding, and rewiring 10\%--50\% of their edges. We embed
these randomized networks into 3D Euclidian space and compute their
AUCs. Noise barely improves the embedding of SF or ER (Figure \ref{fig:rocs_model} (b)
suggesting that the data is unlikely to have a noisy SF or ER
structure; note that with edge deletions and additions the size of the
networks changes affecting the quality of the embedding and thus,
unlike above, we analyze ``randomized'' ER. However, noise has
different effects on the embedding of GEO. Random edge deletions do
not disturb the quality of the geometric embedding, since edge
deletions have little effect on shortest path lengths in GEO
networks. Therefore, AUCs for GEO networks obtained by random edge
deletions are almost the same as AUCs for non-randomized GEO networks
(Figure \ref{fig:rocs_model} (b)). On the other hand, shortest path lengths decrease with
random edge additions and rewirings in GEO networks, resulting in
worse embeddings and lower AUCs (Figure \ref{fig:rocs_model} (b)). Similar is observed for
YH1: random edge deletions do not affect the quality of the embedding,
whereas random edge additions and rewirings result in lower AUCs
(Figure \ref{fig:rocs_model} (b)). The comparable behaviors of GEO and YH1 suggest that
they have similar structures, thus additionally validating our network
model predictions. Moreover, AUC value of 0.87 for GEO with 10\% of
randomly rewired edges is very close to AUC value of 0.89 for YH1
(Figure \ref{fig:rocs_model}). Thus, the structure of the PPI data appears to be
consistent with the structure of a noisy GEO.

\subsection*{Comparison with Other Studies}

Filkov \emph{et al.} use seven network properties to describe a
real-world network and compare it with model networks \cite{Filkov2009}.
In comparison, we use 34 properties,
therefore decreasing the number of
degrees of freedom in which networks being compared can vary.  Also,
the methodology used by Filkov \emph{et al.} is different than
ours.
First, they evaluate the fit of a model to the data by using principal
component analysis of the vector space whose coordinates are the
statistics of the seven network properties that they analyzed.
Second, they evaluate the fit of two scale-free network models to the
data: SF and their new scale-free model of network growth via
sequential attachment of linked node groups.  In comparison, we use three
network models that have very different network structure: ER, SF, and GEO.

Middendorf \emph{et al.} measure the topological structure of a
network by counting the number of random walks of a given length
\cite{Middendorf2004,Middendorf2005} and giving those counts as input
into classifiers.  Random walks are different than graphlets in
several ways.  First, graphlets are induced and random walks are not.
Second, nodes and edges can be repeated in a random walk, while a
graphlet consists of a unique set of nodes and edges.  Middendorf
\emph{et al.}  use two classifiers, SVM \cite{Middendorf2004} and DT
\cite{Middendorf2005} to discriminate different network models. That is,
in each study, they use a single probabilistic method to predict the
best fitting network model for a real-world network. In comparison, in
this study we use five different probabilistic methods, all supporting
GEO as the best-fitting model.  We show that DT and SVM are the least
robust out of the five probabilistic methods that we analyzed (see
Section ``Robustness and Validation'' and Table 3).  Moreover, the
training set of Middendorf \emph{et al.} contains model networks of
the size of the data only and thus it could be biased by the model
properties that are enforced by the chosen network size. In
comparison, we train our probabilistic models on random networks of
different sizes (see Sections ``Techniques'' and ``Methods'') to allow
for a possibility to predict the best fitting network model for any
yet unseen real-world network, independent of its size.  Middendorf
\emph{et al.} consider SF, ER, and small-world networks and identify
SF-based duplication-mutation models as the best-fitting models for
biological networks.  Given that Middendorf \emph{et al.} did not
consider GEO in their studies, and given a low robustness of DT and
SVM that they used, their reported best fit of SF-based models to the
data could be questioned.

\subsection*{Discussion}

\begin{figure*}[!ht]
\begin{center}
\textbf{(a)}\rotatebox{-90}{\resizebox{0.331\textwidth}{!}{\includegraphics{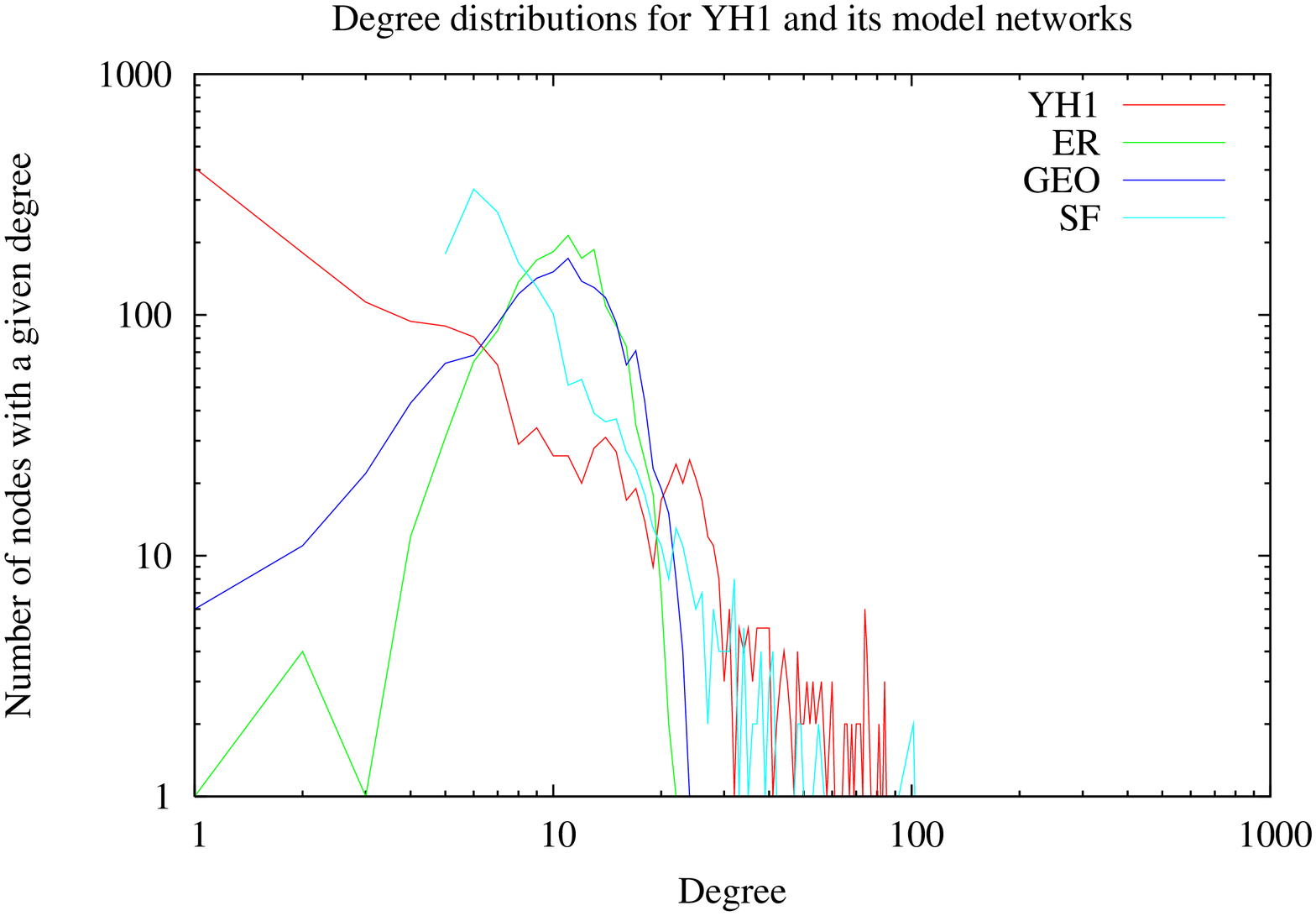}}}
\textbf{(b)}\rotatebox{-90}{\resizebox{0.331\textwidth}{!}{\includegraphics{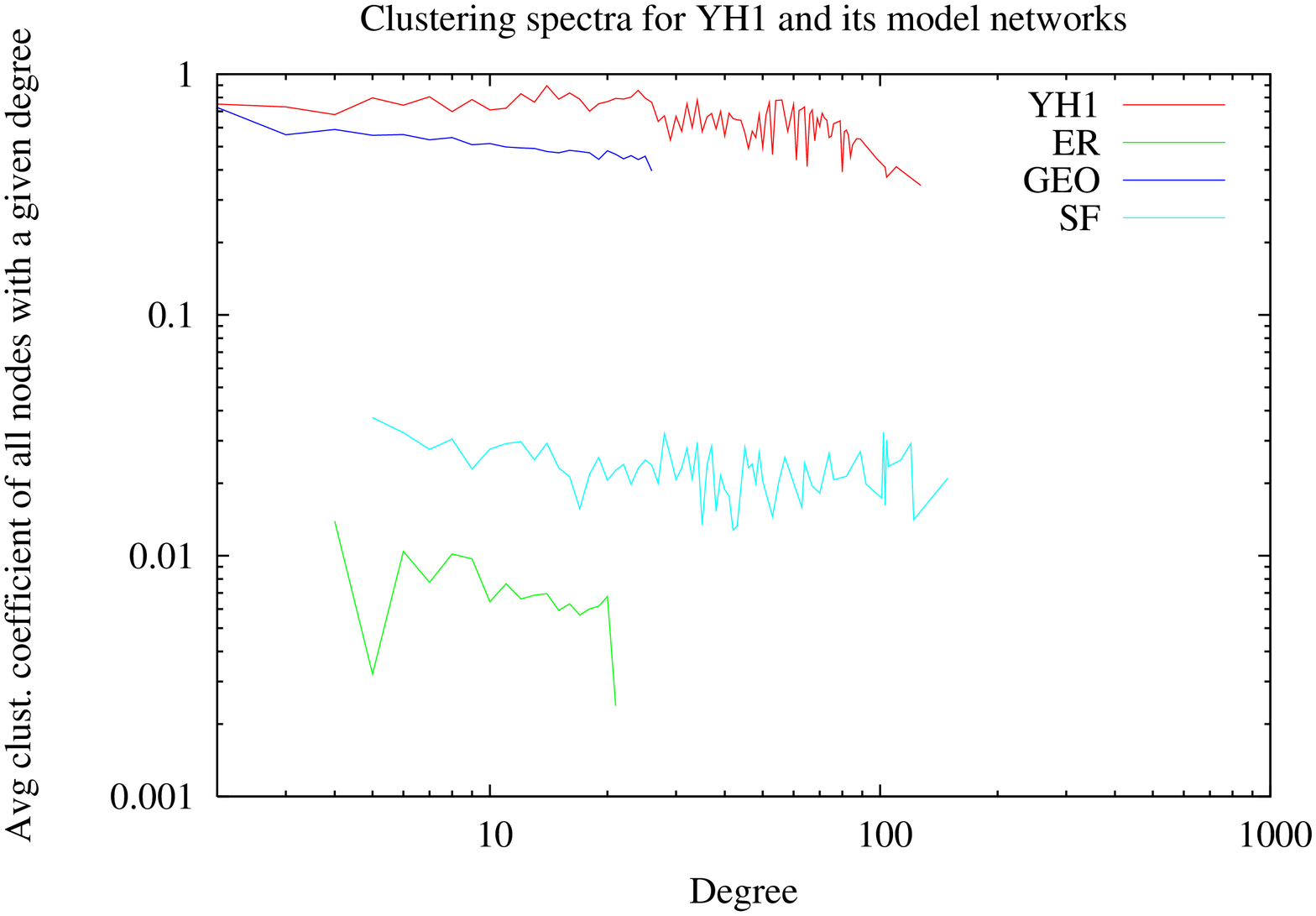}}}
\textbf{(c)}\rotatebox{-90}{\resizebox{0.331\textwidth}{!}{\includegraphics{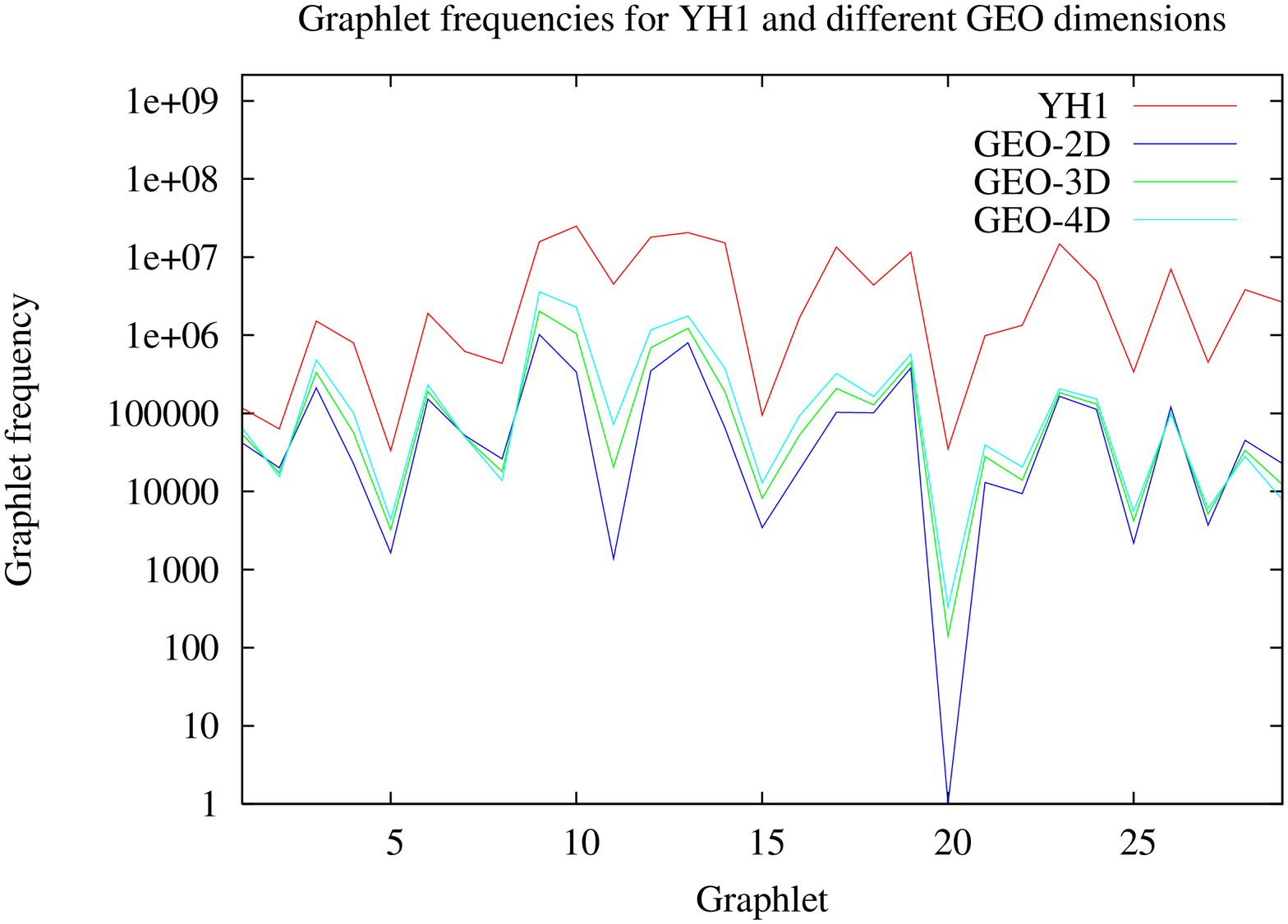}}}
\textbf{(d)}\rotatebox{-90}{\resizebox{0.331\textwidth}{!}{\includegraphics{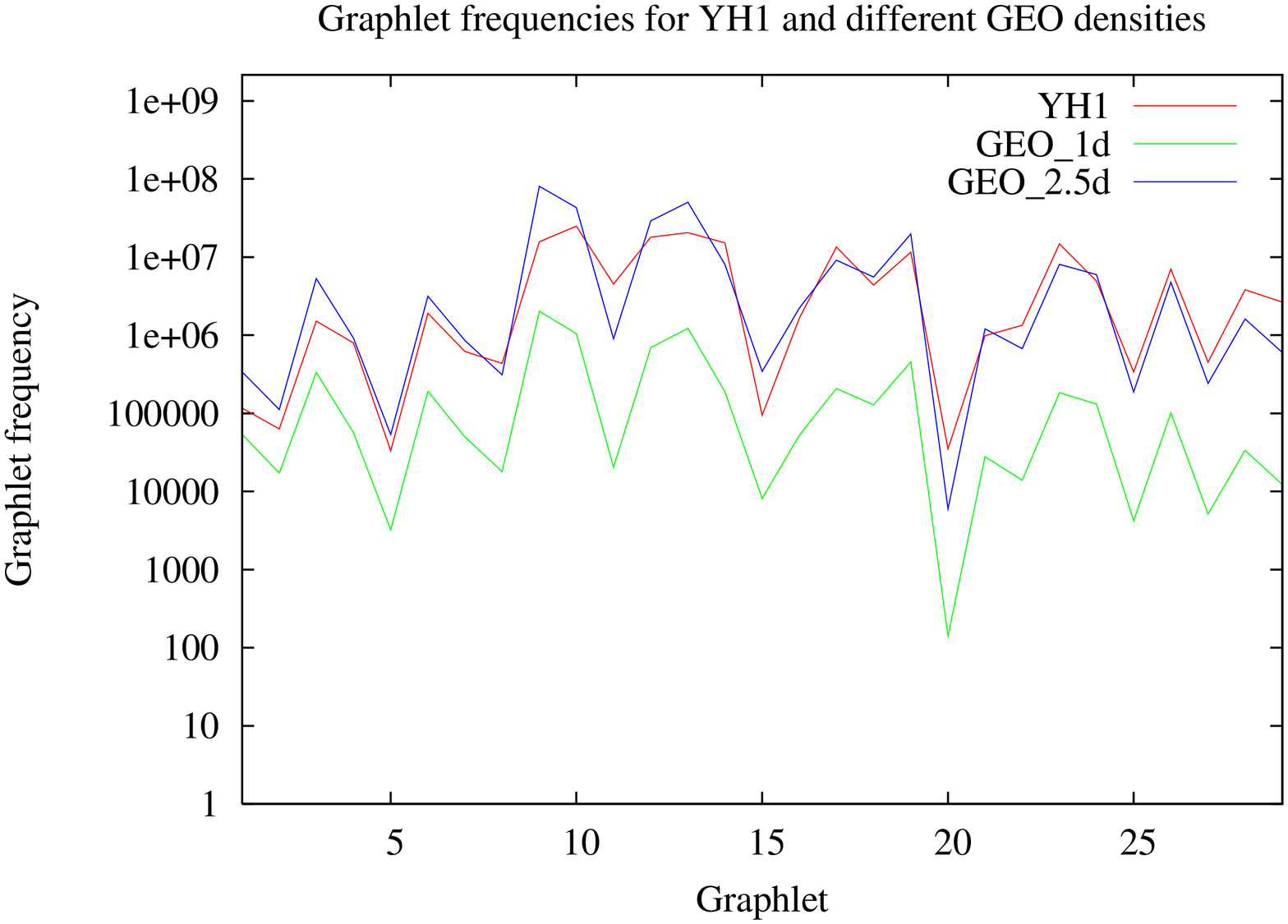}}}
\end{center}
\caption{\textbf{(a)} Degree distributions and \textbf{(b)} clustering spectra for YH1, and an ER, a GEO, and an SF model network of the same size as YH1. \textbf{(c)} Graphlet frequencies for YH1, a 2-dimensional GEO (``GEO-2D''), a 3-dimensional GEO (``GEO-3D''), and a 4-dimensional GEO (``GEO-4D'') network of the same size as YH1. \textbf{(d)} Graphlet frequencies for YH1, a GEO network with the same number of nodes and edges as YH1 (``GEO\_1d''), and a GEO
network with the same number of nodes, but 2.5 times as many edges as YH1 (``GEO\_2.5d'').  On horizontal axes in panels c and d, graphlets are numbered as in Figure \ref{fig:heur} A.}
\label{fig:properties}
\end{figure*}

We further elaborate on the power of integration of different network
properties as opposed to using individual ones to asses the fit of a
network model to the data.  We use our GraphCrunch software package
\cite{GraphCrunch} to evaluate the fit of ER, GEO, and SF models to
all PPI networks described in Data Sets section.  GraphCrunch
evaluates the fit of the models to the data with respect to seven
local and global network properties.  When we evaluate the fit of the
data to the models with respect to each of the seven properties, we
obtain inconclusive results, because each of the properties favors a
different model.  For example, as illustrated in Figure \ref{fig:properties} (a), SF fits
YH1 the best with respect to the degree distribution, but GEO is the
best-fitting network model with respect to the clustering spectrum
(Figure \ref{fig:properties} (b)). This demonstrates the need for a method that finds a
consensus between models suggested by different network properties.
We propose such a method in this study.  Since our method integrates a
variety of network properties, it imposes a large number of
constraints on the networks being compared and reduces the number of
degrees of freedom in which they can differ, thus increasing the
confidence in the fit of a network model.  Inclusion of additional
network properties could further increase the confidence at the
expense of an increased computational complexity.

Although several studies proposed GEO as a well-fitting null model for
PPI networks \cite{Przulj04,Przulj06,HRP08}, a recent
study questioned this \cite{Colak2009}. Note however, that this
conclusion was based on analyzing only one eukaryotic and one
prokaryotic PPI network \cite{Colak2009}, each from DIP \cite{DIP}.
Thus, in the light of low quality and incompleteness of the data from
large databases \cite{Cusick2009} (also see Results section), no
conclusions about the fit of GEO should have been made. The authors
argued that low-dimensional geometric random graphs might not be able
to capture high abundance of dense graphlets and bipartite subgraphs
observed in real-world networks, neglecting two obvious alternatives
for reconciling the differences in the abundance of subgraphs in the
data and in GEO: (1) they based their conclusion on the observation
that bipartite graphlet 20 cannot exist in 2-dimensional (2D) GEO,
even though it exists in 3D GEO as well as in all higher dimensions
(Figure \ref{fig:properties} (c)); and (2) GEO graphs with the same number of nodes, but
$2.5$ times more edges than the data have very similar abundance of
all graphlets as the data (Figure \ref{fig:properties} (d)). These observations are
important since: (1) the optimal dimension for the space of PPI
networks is unknown and finding it is a non-trivial research problem,
but it is highly unlikely that PPI networks exist in a 2D space; and
(2) the density of real-world data will continue to increase
\cite{Stumpf2008,Yu2008}, most likely in accordance with its network
model. Also note that over-abundance of graphlets in the currently
available PPI networks could be an artefact of the ``matrix'' and
``spoke'' models used to determine PPIs in affinity purification
followed by mass spectrometry (AP/MS) pull-down experiments.  In the
matrix model, interactions are defined between all proteins in a
purified complex, clearly resulting in over-abundance of dense
graphlets. In the spoke model, interactions are defined between a bait
and each of its preys, but not between the preys, clearly resulting in
over-abundance of sparse graphlets; an overlap between preys in
different purifications results in over-abundance of complete
bipartite graphs.

Finally, it is possible that PPI networks are not completely geometric
and that another random graph model would provide a better
fit. Additionally, it is possible that different parts of PPI networks
have different structure. The two most commonly used high-throughput PPI detection
methods, AP/MS and Y2H, are fundamentally different: AP/MS detects
mostly stable protein complexes, whereas Y2H detects mostly transient
signalling interactions \cite{Yu2008}. Thus, the two methods examine
different, complementary subspaces within the interactome, resulting
in networks with different topological and biological properties
\cite{Yu2008}. Since proteins within a protein complex are close in
the cell, it is possible that protein complexes have a geometric
structure.  In contrast, transient interactions in signalling pathways
might have a different structure, such as that of bipartite graphs or
scale-free networks.

\section*{Conclusions}
We present an integrative approach for identifying the best-fitting
network model for RIGs and PPI networks. We use five probabilistic
methods to evaluate the fit of a network model to a real-world network
with respect to a series of network properties. All
five probabilistic methods confirm that GEO is the best-fitting model
for RIGs. PPI data sets of high confidence and coverage
are also fitted the best by GEO, while sparser and lower-confidence
PPI networks are fitted the best by SF or ER. By testing the
robustness of probabilistic methods to noise, we demonstrate that our
approach is unlikely to predict a real-world network as GEO if it had
a noisy SF structure, independent of the probabilistic method or
noise level. On the other hand, it could classify a real-world network
as SF or ER if it had a noisy GEO structure, depending on the
probabilistic method and noise level. Together, these results suggest that the
structure of PPI networks is the most consistent with the
structure of noisy GEO.
%
Since networks have been used to model real-world phenomena in various
domains, it would be interesting to apply our method to other types of
real-world networks, such as technological or social ones.

\section*{Methods}

\subsection*{Network fingerprint}

We summarize the structure of a complex network by the notion of the
``network fingerprint'' (of just ``fingerprint,'' for brevity). We
define the \emph{fingerprint} to be a 34-dimensional vector whose
coordinates contain the following network properties: the average
degree, average clustering coefficient, average diameter, and
frequencies of the appearance of all 31 1-5-node graphlets. The
\emph{degree} of a node is the number of edges incident to the node;
the \emph{average degree} of a network is the average of degrees over
all nodes in the network.  The \emph{clustering coefficient} of a node
is defined as the probability that two neighbors of the node are
themselves connected. The average of clustering coefficients over all
nodes in a network is the \emph{average clustering coefficient} of the
network.  The smallest number of links that have to be traversed in a
network to get from one node to another is called the \emph{distance}
between the two nodes and a path through the network that achieves
this distance is called the \emph{shortest path} between the nodes;
the average of shortest path lengths over all pairs of nodes in a
network is called the \emph{average network diameter}.
\emph{Graphlets} are small connected non-isomorphic induced subgraphs
of a large network \cite{Przulj04}; we count the occurrences of the
only 1-node graphlet, a node, the only 2-node graphlet, an edge, and
all 29 3-5-node graphlets (shown in Figure 1(a)).  Because different
coordinates of a network fingerprint can differ by several orders of
magnitude, we normalize each coordinate to avoid domination of
coordinates having large values. We normalize the $i^{th}$ coordinate
$x_i$ of the network fingerprint \emph{x} as $log(x_{i} + 1)$, for
$i=1,...,34$; we add 1 to $x_{i}$ to avoid the logarithm function to
go to infinity when $x_{i}=0$.

\subsection*{Random network models}

We consider three random network models: Erd\"{o}s-R\'{e}nyi (ER)
random graphs \cite{ErdosRenyi59}, scale-free Barab\'{a}si-Albert (SF)
networks \cite{Barabasi99}, and geometric (GEO) random graphs
\cite{Penrose03}. In \emph{Erd\"{o}s-R\'{e}nyi random graphs}, edges
between pairs of nodes are distributed uniformly at random with the
same probability $p$ \cite{ErdosRenyi59}.  \emph{Scale-free networks}
are networks that have power-law degree distributions. The version of
SF networks that we use are generated by Barab\'{a}si-Albert
peripheral attachment method \cite{Barabasi99}, in which newly added
nodes preferentially attach to existing nodes with probabilities
proportional to their degrees. In \emph{geometric random graphs},
nodes correspond to uniformly distributed points in a metric space and
edges are created between pairs of nodes if the corresponding points
are close enough in the metric space according to some distance norm
\cite{Penrose03}. We construct geometric random graphs by using
3-dimensional Euclidean boxes and the Euclidean distance norm
\cite{Przulj04}.

For each of the three random network models, we generate 10 instances
of random networks per model. We generate random networks of different
sizes, both in terms of the number of nodes ($n$) and the number of
edges ($m$).  We use the following $28$ values for $n$: 100, 200, 300,
400, 500, 600, 700, 800, 900, 1,000, 1,100, 1,200, 1,300, 1,400,
1,600, 2,100, 2,600, 3,100, 3,600, 4,100, 4,600, 5,000, 6,000, 7,000,
8,000, 9,000, 10,000, and 11,000. For each of the $26$ values of $n$
below 10,000, we vary $k=m/n$ from 1 to 10, in increments of 1. Due to
the increase in computational complexity with the increase in the
number of nodes and edges, for the $2$ largest values of $n$,
$n=10,000$ and $n=11,000$, we only use $k = 1,...,7$.  Thus, we
analyze $26 \times 10 + 2 \times 7 = 274 $ different network sizes. In
total, for the 3 network models, 10 random network instances per
model, and 274 network sizes, we create $3 \times 10 \times 274 =
8,220$ model networks.

\subsection*{Probabilistic methods}

We use five well-known probabilistic methods: backpropagation method
(BP), probabilistic neural networks (PNN), decision tree (DT),
multinomial na\"{i}ve Bayes classifier (MNB), and support vector
machine (SVM).

Both \textbf{BP} \cite{MatBpg} and \textbf{PNN} \cite{MatPnn} are
based on artificial neural networks (ANNs).  ANNs are simplified
mathematical models of biological nervous systems built of processing
units called neurons. Neurons in ANNs have many input signals and they
produce one output signal. They are organized into the following
layers: the input layer, one or more hidden layers, and the output
layer.  Neurons in the input layer do not perform any processing;
instead, they only distribute the input data to all neurons in the
first hidden layer. The number of hidden layers depends on
implementation of an ANN. We use the standard implementations of BP
and PNN from Neural-Network Toolbox in
Matlab\footnote{http://www.mathworks.com/access/helpdesk/help/toolbox/nnet/}. For
the completeness of the manuscript, we briefly outline them below.

In our implementation of \textbf{BP} \cite{MatBpg}, the input layer
consists of 34 neurons corresponding to the 34 coordinates of the
network fingerprint input vector.  To match the length of our input
vector, we implement one hidden layer with 15 neurons; varying the
number of neurons in the hidden layer between 10 and 20 had marginal
effect on the results. The output layer contains three neurons,
according to the ``1-of-N encoding of the output classes'' principle
\cite{Stegemann1999}: the number of neurons in the output layer
matches the number of possible ``output classes,'' i.e., random
network models (ER, SF, and GEO). Thus, for a given output class, the
neuron corresponding to the class is set to 1, whereas the remaining
two neurons are set to -1.  After BP computes the values on the three
output neurons for an input vector, it classifies the input into an
output class that corresponds to the neuron with the largest value.

All neurons in the input layer are connected with all neurons in the
hidden layer. Similarly, all neurons in the hidden layer are connected
with all neurons in the output layer. All of these connections are
weighted. Each neuron in the hidden and the output layer produces
output by applying a non-linear ``transfer function'' to calculate a
weighted sum of its inputs. We use \emph{logsig} and \emph{tansig}
transfer functions in the hidden layer and the output layer,
respectively. Initially, all weights are assigned randomly. Weights
are adjusted gradually trough a training (learning) process: BP keeps
adjusting the weights until the error between the value of each output
neuron and its desired value (i.e., the value of the class that the
input that we are training BP on belongs to) is $\leq 10^{-5}$.  We
use \emph{trainscg} ``training function'' and set the learning rate to
0.01.  Given these parameters, BP is successfully trained on the
training set (defined in Techniques section above) in 588 epochs.

\textbf{PNN} that we use consists of the radial basis layer and the
competitive layer. The radial basis layer further consists of the
input and pattern sublayers.  Similarly, the competitive layer
consists of the summation and the output sublayers. The number of
neurons in the input sublayer corresponds to the 34 dimensions of the
input vector. The pattern sublayer consists of three pools of
``pattern'' neurons, where each pool corresponds to one of the three
output classes.  The number of neurons in each pool is determined as
follows. As each network fingerprint from the training set is provided
as input vector into PNN during the training process, a new neuron is
added to the pool that corresponds to the output class (i.e., network
model) of the input vector.  After the training phase, when an input
vector is presented to the trained PNN, the pattern sublayer computes
how close the input vector is to each of the vectors from the training
set in each pool. This information is sent to the summation
sublayer. The summation sublayer consists of three neurons, where each
neuron corresponds to one of the three output classes. Input into each
neuron in the summation sublayer is the collection of outputs from the
corresponding pool in the pattern sublayer. The output of each
summation sublayer neuron is a weighted sum of all its inputs. Each of
the three sums represents the probability that the input vector
belongs to the corresponding class. Given these probabilities, the
output sublayer, consisting of a single neuron, outputs the class
having the highest probability.

We use a standard implementation of \textbf{DT} \cite{MatDt,ChBook}
from Statistics Toolbox in
Matlab\footnote{http://www.mathworks.com/access/helpdesk/help/toolbox/stats/}. Interior
nodes in the decision tree are queries on certain attributes; in our
case, attributes are the coordinates of the fingerprint vector. Each
leaf in the tree corresponds to one of the three output classes.
Branches in the tree represent conjunctions of attributes that lead to
classification into the output classes. DT recursively splits the
training set of input vectors into subsets based on the values of
their coordinates; this corresponds to branching in the tree. DT
continues to do so until the training input vectors are assigned to
their correct classes.

We use a standard implementation of \textbf{MNB} \cite{ChBook} from
WEKA \cite{RWeka}, a publicly available collection of machine learning
algorithms for data mining. MNB classifies the input data based on the
Bayes' rule by selecting a class that maximizes the posterior
probability of the class, given the training set. MNB does not use the
assumption of a na\"{i}ve Bayes classifier, that all data attributes
are independent of each other.

We use a standard implementation of \textbf{SVM}
\cite{ChBook,Vapnik1999} from WEKA.  SVM maps our 34-dimensional input
vectors into a high dimensional space; the space dimension is
automatically determined by WEKA. During the training phase, SVM finds
an optimized data division within this space by constructing a
hyperplane that optimally separates the data into two classes; since
there are many hyperplanes that might classify the data, the
hyperplane is chosen so that the distance from the hyperplane to the
nearest data point is maximized. We generalize this binary
classification to the multiclass classification, with three classes
corresponding to the three random network models. We do so by using
three binary ``one-versus-all'' SVMs: for each of the three classes,
its corresponding SVM either classifies the input data as belonging to
the class (``positive classification''), or not belonging to the class
(``negative classification'') \cite{ChBook,Vapnik1999}. Each of these
three binary SVMs produces an output function that gives a relatively
large value for a positive classification and a relatively small value
for a negative classification. The input data is classified into the
class with the highest value of the output function.

\section*{Authors contributions}

VM participated in the design of the study, carried out most of the
experiments, analyzed the results, and helped write the paper.
TM participated in the design of the study, helped run the
experiments, analyzed the results, and wrote the paper.
NP conceived of the study and participated in its design and
coordination and helped write the manuscript.
All authors read and approved the final manuscript.

\section*{Acknowledgements}
%
%
This project was supported by the NSF CAREER IIS-0644424
grant.




\end{document}